\documentclass[12pt]{article}

\usepackage{amssymb,epsfig}

\usepackage{psfrag}

\psfrag{phi(k)}{{\footnotesize $\phi_+(k,\mu=1$~GeV)}}
\psfrag{phi(t)}{{\footnotesize $\varphi_+(\tau,\mu=1$~GeV)}}

\psfrag{(a)}{{\footnotesize $(a)$}}
\psfrag{(b)}{{\footnotesize $(b)$}}
\psfrag{(c)}{{\footnotesize $(c)$}}
\psfrag{(d)}{{\footnotesize $(d)$}}
\psfrag{(e)}{{\footnotesize $(e)$}}
\psfrag{(f)}{{\footnotesize $(f)$}}
\psfrag{(g)}{{\footnotesize $(g)$}}
\psfrag{(h)}{{\footnotesize $(h)$}}

\psfrag{tn}{{\footnotesize $tn$}}
\psfrag{x}{{\footnotesize $x$}}

\psfrag{k, GeV}{{\footnotesize $k$, GeV}}
\psfrag{t, GeV}{{\footnotesize $\tau$, GeV$^{-1}$}}

\psfrag{I2}{{$|{\cal I}|^2$}}
\psfrag{zz}{{$z$}}
\psfrag{zeta}{{$\zeta$}}
\psfrag{TH}{\footnotesize $T_H(z',x_1)$}

\newcommand{\beq}[1]{
\begin{equation}\label{#1}}
\newcommand{\eeq}{\end{equation}}
\newcommand{\bea}[1]{
\begin{eqnarray}\label{#1}}
\newcommand{\eea}{\end{eqnarray}}
\newcommand\re[1]{(\ref{#1})}
\def \e  {\mathop{\rm e}\nolimits}

\begin{document}

\begin{titlepage}

\begin{flushright}
\begin{tabular}{l}
LPT--Orsay--03--63\\
 hep-ph/0309330
\end{tabular}
\end{flushright}
\vspace{1.5cm}

\begin{center}
{\LARGE \bf
The B-Meson Distribution Amplitude in QCD
}
\vspace{1cm}

{\sc V.M.~Braun}${}^1$, {\sc D.Yu.~Ivanov}${}^{2}$ and {\sc
G.P.~Korchemsky}${}^3$
\\[0.5cm]
\vspace*{0.1cm} ${}^1${\it
   Institut f\"ur Theoretische Physik, Universit\"at
   Regensburg, \\ D-93040 Regensburg, Germany
                       } \\[0.2cm]
\vspace*{0.1cm} ${}^2${\it Institute of Mathematics, 630090 Novosibirsk, Russia
                       } \\[0.2cm]
\vspace*{0.1cm} ${}^3${\it Laboratoire de Physique Th\'eorique%
\footnote{Unite Mixte de Recherche du CNRS (UMR 8627)},
Universit\'e de Paris XI, \\
91405 Orsay C\'edex, France} \\[1.0cm]

\vspace{0.6cm}
\vskip1.2cm
{\bf Abstract:\\[10pt]} \parbox[t]{\textwidth}{
 The B-meson distribution amplitude is calculated using QCD sum rules.
 In particular we obtain an estimate for the integral relevant to
exclusive B-decays $\lambda_B = 460\pm 110$~MeV at the scale 1 GeV.
A simple QCD-motivated parametrization of the distribution amplitude
is suggested.
}
  \vskip1cm
\end{center}

\end{titlepage}


\newpage

\section{Introduction}
\setcounter{equation}{0}

The B-meson distribution amplitude was introduced in \cite{SHB90} as the direct
analogue of light-cone distribution amplitudes of light mesons
\cite{earlyCZ,earlyBL,earlyER} in an attempt to describe generic exclusive
B-decays  by the contribution of the hard gluon exchange. Since then,
considerable effort has been invested to understand the QCD dynamics of heavy
meson decays in the heavy quark limit. The radiative decay $B\to\gamma e\nu$
provides one with the simplest example of such processes \cite{KPY}. This form
factor can be calculated in terms of the B-meson distribution amplitude to
one-loop accuracy \cite{DS03} and arguments have been given that the
corresponding factorization formula is valid to all orders in the strong coupling
\cite{proofsBgamma}. Similar QCD factorization formulas have also been proposed
for the related processes $B\to\gamma\gamma$ and $B\to\gamma\ell^+\ell^-$
\cite{BB02,DS03a}. For weak decays involving energetic light hadrons in the final
states the QCD factorization is more complex since one must isolate the end-point
soft contributions in terms of additive contributions. This is a hot topic, see
e.g. \cite{friedhof}, and the results have been encouraging although, as it has
been repeatedly pointed out \cite{ABS94,Ball03}, the $1/m_b$ corrections to
heavy-to-light exclusive decays are most likely large and require quantitative
treatment.

The B-meson distribution amplitude plays the central role in all known
factorization formulas, but, surprisingly, received relatively little attention
in the past. In the present work we use QCD sum rules to present a realistic
model for the B-meson distribution amplitude, consistent with all QCD
constraints. We also take this opportunity to clarify its theoretical status on
which there has been certain confusion. The approach that we take in this paper
is
inspired by
the classical work by Chernyak and Zhitnitsky on the QCD sum rule analysis of
the distribution amplitudes of light mesons \cite{CZ}. In particular we argue
that the relevant matrix element of a bilocal quark-antiquark operator can be
calculated by the QCD-corrected expansion at {\it imaginary} light-cone
distances. We find that the nonperturbative corrections remain under control and
present quantitative estimates for the distribution amplitude and its first
inverse moment which enters decay form factors at tree level. Our results can be
considered as an extension of an earlier QCD sum rule calculation by Grozin and
Neubert \cite{GN96} (see also \cite{BK03}) with the main difference being that we
calculate NLO radiative corrections to the sum rule. This is important since the
true analytic structure of the distribution amplitude is only seen at this level.

The presentation is organized as follows. Sec.~2 is introductory and contains the
necessary definitions. A simple model for the distribution is obtained in Sec.~3
using QCD perturbation theory and duality. The complete sum rule is constructed
in Sec.~4 where we discuss the structure of nonperturbative corrections. This
Section also contains our main results, the summary  and conclusions.

\section{Definitions}

{}Following Ref.~\cite{LN03} we define the B-meson distribution
amplitude as the renormalized matrix element of the bilocal operator
built of an effective heavy quark field $h_v(0)$ and a light antiquark
$\bar q(tn)$ at a light-like separation:
\beq{defDA}
  \langle 0|\Big[\bar q(tn)\!\not\!n
       [tn,0]\Gamma  h_v(0)\Big]_R|\bar B(v)\rangle =
  -\frac{i}{2}F(\mu)\,\mbox{\rm Tr}\left[\gamma_5\!\not\!n\Gamma
   P_+\right]\,\Phi_+(t,\mu)\,
\eeq
with
\beq{WL}
 {}[tn,0] \equiv {\rm Pexp}\left[ig\int_0^t\!du\,n_\mu A^\mu(utn)\right].
\eeq
Here $v_\mu$ is the heavy quark velocity,
$n_\mu$ is the light-like vector, $n^2=0$, such that $n\cdot v=1$,
$P_+=\frac12(1+\not\!v)$ is the projector on upper components of the
heavy quark spinor, $\Gamma$ stands for an arbitrary Dirac structure,
$|\bar B(v)\rangle$ is the $\bar B$-meson state in the heavy quark
effective theory (HQET) and $F(\mu)$ is the decay constant in HQET,
which is related to the physical B-meson decay constant to one-loop
accuracy as
\beq{defF}
   f_B\sqrt{m_B} = F(\mu)\left[1+\frac{C_F\alpha_s}{4\pi}
   \left(3\ln\frac{m_b}{\mu}-2\right)+\ldots\right].
\eeq
The notation $[\ldots]_R$ in \re{defDA} stands for the renormalization in a
MS-like scheme and $\mu$ here and below refers to the $\overline{\textrm{MS}}$
normalization scale.

The invariant function $\Phi_+(t,\mu)$ where $t$ is a real number
defines what is usually called the leading twist
B-meson distribution amplitude in position space, in contrast to the
amplitude $\Phi_-(t,\mu)$ which involves a different light-cone projector
--- $\not\!\bar n$ instead of $\not\!n$ --- in between the quarks;
here $\bar n^2 =0, \bar n \cdot n = 2$. This name is not exact
since the translation symmetry of the theory is broken by presence
of the effective heavy quark field and hence neither geometrical nor
collinear twist are defined. In the present paper we only consider
the distribution $\Phi_+(t,\mu)$ and its Fourier transform
\bea{Fourier}
     \phi_+(k,\mu)&=&\frac{1}{2\pi}\int_{-\infty}^{\infty}\!dt \, \e^{ikt}\Phi_+(t-i0,\mu)\,,
\nonumber\\
     \Phi_+(t,\mu)&=&\int_0^\infty\!dk \, \e^{-ikt}\phi_+(k,\mu)\,,
\eea
where in the first equation the integration contour goes below the singularities
of $\Phi_+(t,\mu)$ that are located in the upper-half plane.

The scale dependence of the distribution amplitude is driven by the
renormalization of the corresponding nonlocal operator
$O_+(t)= \bar q(tn)\not\!\!n\,[tn,0]\Gamma\, h_v(0)$. The corresponding
$Z$-factor was computed in \cite{LN03} to one-loop order. In momentum
space, the result reads
\beq{defZ}
     O_+^{\rm ren} (k,\mu) = \int dk'\, Z_+(k,k';\mu)\,O_+^{\rm bare}(k')\,,
\eeq
where
\bea{Zmom}
   Z_+(k,k';\mu) &=& \delta(k-k')+ \frac{\alpha_sC_F}{4\pi}
Z^{(1)}_+(k,k';\mu)+\ldots\,,
\nonumber\\
  Z^{(1)}_+(k,k';\mu) &=&\left(\frac{4}{\hat\epsilon^2}
  +\frac{4}{\hat\epsilon}\ln\frac{\mu}{k}-\frac{5}{\hat\epsilon}\right)
  \delta(k-k')-\frac{4}{\hat\epsilon}
  \left[\frac{k}{k'}\frac{\theta(k'-k)}{k'-k}+\frac{\theta(k-k')}{k-k'}
  \right]_+
\eea
with $d=4-\epsilon$ and the standard notation
$2/\hat\epsilon = 2/\epsilon-\gamma_E+\ln 4\pi$. Here $[\ldots]_+$ is the
usual ``plus''-distribution.

In order to understand the meaning of this result it is instructive
to consider operator renormalization in position space. The
corresponding to \re{Zmom} one-loop expression is
\bea{Zcoord}
 O_+^{\rm ren} (t,\mu) &=& O_+^{\rm bare}(t)+
 \frac{\alpha_sC_F}{4\pi}\left\{\left(\frac{4}{\hat\epsilon^2} +
 \frac{4}{\hat\epsilon}\ln(i t\mu )\right)O_+^{\rm bare}(t)
\right.\nonumber\\&&{}\left.
 - \frac{4}{\hat\epsilon}\int_0^1\!du\,\frac{u}{1-u}
    \left[O_+^{\rm bare}(ut)-O_+^{\rm bare}(t)\right]
  - \frac{1}{\hat\epsilon}O_+^{\rm bare}(t)\right\},
\eea
%
\begin{figure}[t]
\centerline{\epsfysize3.7cm\epsffile{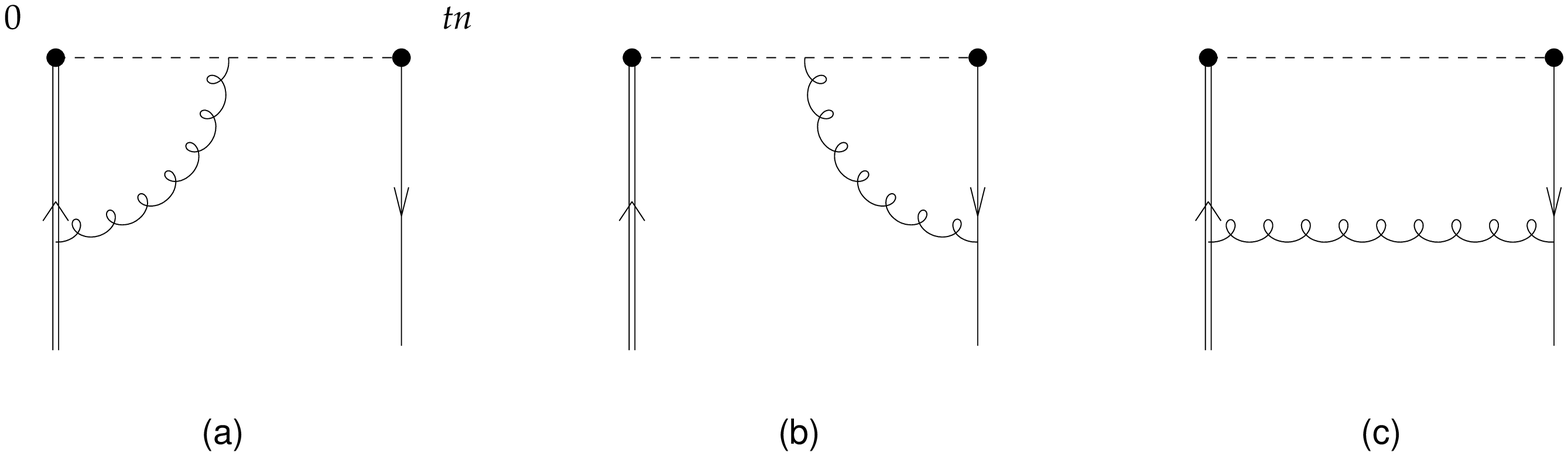}}
\caption[]{\small
One-loop renormalization of the nonlocal light-cone operator built of
one light and one effective heavy quark field (double line).
The dashed line indicates the gluon Wilson line insertion
in between the quark fields.
 }
\label{fig:1}
\end{figure}
%
where the first two terms in curly brackets correspond to vertex-type corrections
shown in Fig.~1a and Fig.~1b, respectively, (in Feynman gauge) and the third term
takes into account the quark field renormalization: $q^{\rm ren} =
Z_q^{-1/2}q^{\rm bare}$ and $h_v^{\rm ren} = Z_h^{-1/2}h_v^{\rm bare}$ with $Z_q
= 1-(2/\hat\epsilon) (\alpha_sC_F/4\pi)$, $Z_h = 1+(4/\hat\epsilon)
(\alpha_sC_F/4\pi)$ \cite{Neubert93}. The exchange diagram in Fig.~1c is
UV-finite and does not contribute \cite{GN96,LN03}.

Note the following property: renormalization of the nonlocal light-cone operator
$O_+(t)$  \re{Zcoord} is quasilocal: it only gets mixed with itself and with
operators with smaller light-cone separation. In fact the heavy quark vertex
correction in Fig.~1a corresponds to a multiplicative (cusp) renormalization in
coordinate space \cite{Korchemskaya} while the light quark vertex correction is
identical to the similar contribution to the light-quark-antiquark nonlocal
operators \cite{BB89}. For light quarks, this property of quasilocality
guarantees existence of the Wilson short distance operator product expansion
(OPE) since it implies that the ``size'' of the operator is not altered by the
renormalization. In the present case, however, the local OPE does not exist
because of the term $\ln(it\mu)$ which is non-analytic at $t\to0$. It is easy to
see that this contribution arizes from the term $\sim (\mu
t)^\epsilon/\epsilon^2$ in the dimensionally regularized diagram in Fig.~1a so
that the answer for this diagram depends on the order of limits $t\to0$ and
$4-d=\epsilon\to0$. We conclude that renormalization of the nonlocal light-cone
operator built of one light and one effective heavy quark field does not commute
with the short-distance expansion. In particular
\beq{noOPE}
 [\bar q(tn)\!\not\!n\,[tn,0]\Gamma\,h_v(0)]_R \,\not=\, \sum_{p=0}^\infty
   \frac{t^p}{p!}[\bar q(0)(\stackrel{\leftarrow}{D}\cdot n)^p h_v(0)]_R,
\eeq
and the equality does not hold even as an asymptotic expansion.
As a consequence,
non-negative moments of the B-meson distribution amplitude $\int\! dk\, k^p
\phi_+(k,\mu)$ for $p=0,1,2,\ldots$ are not related to matrix elements of local
operators and in fact do not exist: It is easy to see that the logarithmic
singularity of the amplitude in position space $\Phi_+(t,\mu)\sim \ln(it)$ for
$t\to0$ implies that the Fourier integral \re{Fourier} is logarithmically
divergent at $k\to \infty$, that is $\phi_+(k,\mu) \sim 1/k $ for $k\gg\mu$, in
agreement with the analysis in \cite{LN03}.
The analysis of moments $\int\! dk\,
k^p \phi_+(k,\mu)$ in \cite{GN96,KKQT01} tacitly assumes a different definition
of the B-meson distribution amplitude, such that the nonlocal
light-cone operator is defined as the generating function for
renormalized local operators on the r.h.s. of \re{noOPE}. This implies
e.g. that power divergences are subtracted.
This is a different object which, most likely, does not satisfy any
simple renormalization group equation and  has no obvious relation
to exclusive B-decays.

\section{Sum Rules: Perturbation Theory}

Aim of the present study is to suggest a realistic model of the B-meson
distribution amplitude that would be consistent with all QCD
constraints. To this end we evaluate the necessary B-meson matrix
elements using the standard QCD sum rule approach \cite{SVZ79}.
In this section we set up the framework and present intermediate results
that only include perturbation theory contributions and the assumption
of duality. The complete treatment including nonperturbative corrections
is presented in the next section.

To derive the sum rules we consider the following correlation functions in
HQET:
\beq{Pi}
   i\int d^4x\, \e^{-i\omega (vx)}\langle0|
   {\rm T}\{\bar q(0)\Gamma_1h_v(0)\bar h_v(x)\Gamma_2q(x)\}|0\rangle =
   -\frac12{\rm Tr}\left[\Gamma_1P_+\Gamma_2\right]\Pi(\omega)
\eeq
and
\beq{T}
   i\int d^4x\, \e^{-i\omega (vx)}\langle0|
   {\rm T}\{\bar q(tn)\!\not\!n\,\Gamma_1[tn,0]h_v(0)
    \bar h_v(x)\Gamma_2q(x)\}|0\rangle =
    -\frac12{\rm Tr}\left[\!\not\!n\,\Gamma_1P_+\Gamma_2\right]T(t,\omega)\,.
\eeq
The correlation function $\Pi(\omega)$ has a pole at $\omega =
\bar\Lambda$ where $\bar\Lambda=m_B-m_b$ is the usual HQET parameter,
and the residue at this pole is proportional to the HQET decay constant
$F(\mu)$:
\beq{lhs2}
   \Pi(\omega) = \frac12 F^2(\mu)\,\frac{1}{\bar \Lambda-\omega}+
   \mbox{\rm higher resonances and continuum}.
\eeq
Similarly,
\beq{lhs3}
   T(t,\omega) = \frac12 F^2(\mu)\,\frac{1}{\bar \Lambda-\omega}
   \int_0^\infty dk\,\e^{-ikt}\phi_+(k,\mu)
+\ldots
\eeq
On the other hand, both correlation functions can be calculated in QCD
at negative values of $\omega$ of the order of 1 GeV in perturbation
theory and taking into account nonperturbative effects induced by
vacuum condensates \cite{SVZ79}. Matching the two representations
one obtains a sum rule. There are two technical details: First, one
makes an assumption that contributions of the continuum and of higher
resonances can be taken into account by the restriction to the
so-called duality region $0<s<\omega_0$ in the dispersion representation
for the correlation functions, e.g.
\beq{duality}
    \Pi(w) = \int_0^\infty \frac{ds}{s-\omega}\, \rho_\Pi(s)
    \to \int_0^{\omega_0} \frac{ds}{s-\omega}\, \rho_\Pi(s)\,,
\eeq
where $\rho_\Pi(s)$ is the corresponding spectral density. The numerical
value for the parameter $\omega_0$ (called continuum threshold) is
usually taken to be in the interval $0.8-1.0$~GeV
\cite{BBBD92,Neubert92,BB94,Neubert93}. Second, one makes the so-called
Borel transformation
\beq{Borel}
  \int_0^{\omega_0} \frac{ds}{s-\omega}\, \rho_\Pi(s) \to
  \int_0^{\omega_0} {ds}\, \e^{-s/M}\, \rho_\Pi(s)
\eeq
introducing the variable $M$ (Borel parameter) instead of the energy $\omega$ in
order to suppress higher-order nonperturbative corrections and minimize the
dependence on the continuum model. The resulting sum rule for the correlation
function $\Pi(w)$ is well known \cite{BBBD92,Neubert92}:
\bea{SR2}
  \frac{1}{2}F^2(\mu)\e^{-\bar\Lambda/M} &=&
   \frac{N_c}{2\pi^2}\int_0^{\omega_0}ds\, s^2\, \e^{-s/M}
   \left[1+\frac{\alpha_s}{\pi}\left(
   \frac{17}{3}+\frac{4\pi^2}{9}-2\ln\frac{2s}{\mu}\right)\right]
\nonumber\\&&{}
 -\frac12\langle\bar q q\rangle
 \left[1+\frac{2\alpha_s}{\pi}-\frac{m_0^2}{16M^2}\right].
\eea
Here $\alpha_s=\alpha_s(\mu)$, $\langle\bar q q\rangle \simeq -(240$~MeV$)^3$ is
the quark condensate and $m_0^2$ is the ratio of the quark-gluon and quark
condensates $m_0^2= \langle \bar q g (\sigma G) q\rangle /\langle\bar q q\rangle
\simeq 0.8$~GeV$^2$. The given numbers correspond to the renormalization scale
$\mu=1$~GeV. With the choice $\bar\Lambda =0.4-0.5$~GeV and $w_0 = 0.8-1.0$~GeV
the sum rule in \re{SR2} is satisfied for a wide range of values of the Borel
parameter $0.3$~GeV$< M < \infty$ and is used \cite{BBBD92,Neubert92} to
determine the B-meson decay constant $F(\mu)$ in the heavy quark limit. In the
numerical estimates in this paper we will take the ``window'' $0.3~\textrm{GeV} <
M < 0.6~\textrm{GeV}$ in which the matching is done \cite{BB94} and use the value
$\alpha_s(1$~GeV$)=0.5$ ($\Lambda_{\rm QCD}^{(3){\rm NLO}}\simeq 360$~MeV) which
is consistent with the world average.

Our task in this work is to derive the similar sum rule for the
correlation function $T(t,\omega)$ defined in \re{T}. The
perturbative contributions are shown in Fig.~2.
%
\begin{figure}[t]
\centerline{\epsfxsize12.0cm\epsffile{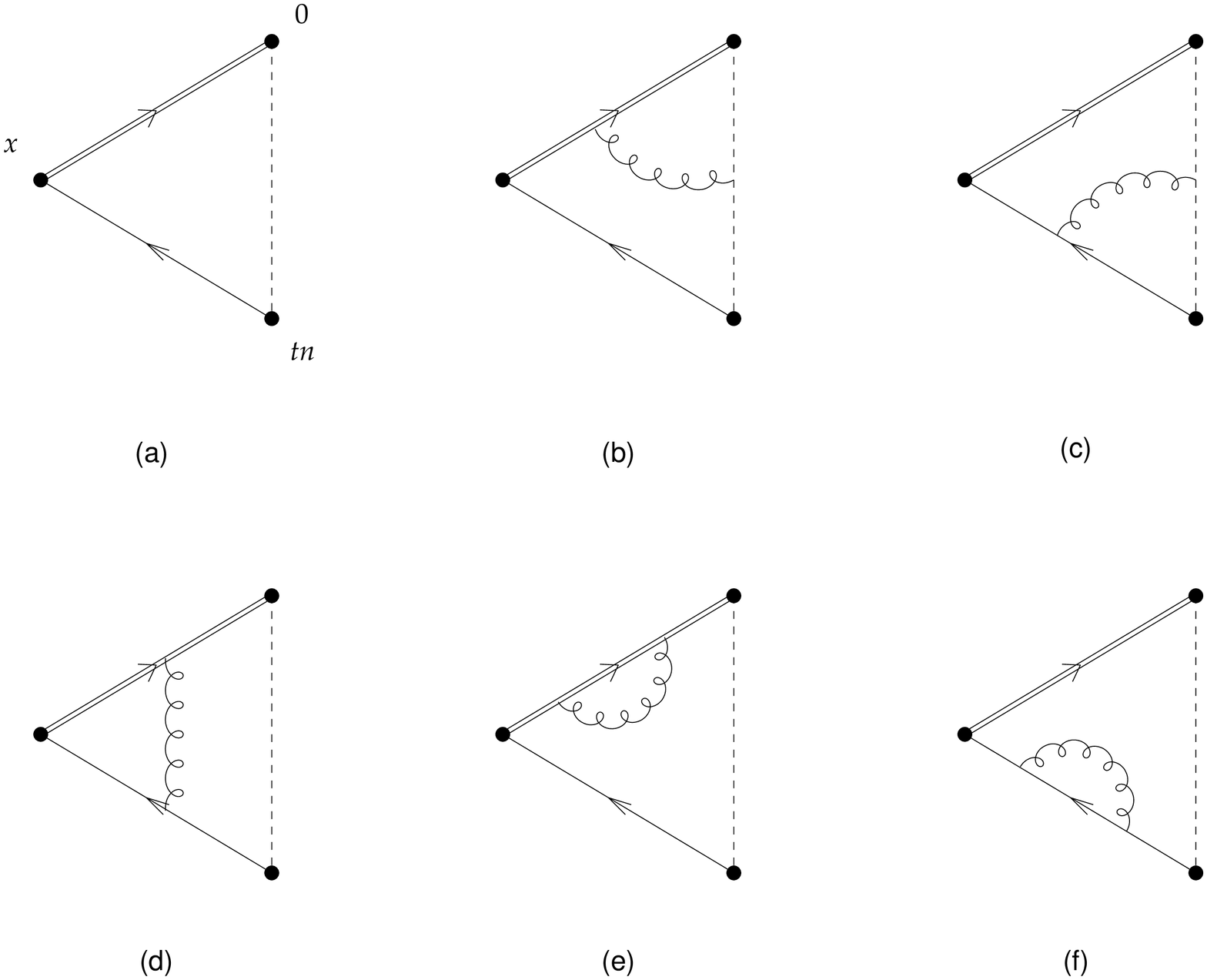}}
\caption[]{\small
Correlation function \re{T} in QCD perturbation theory to first order.
 }
\label{fig:2}
\end{figure}
%
The corresponding sum rule reads, so far without nonperturbative
corrections:
\bea{SR3a}
  \frac{1}{2}F^2(\mu)\e^{-\bar\Lambda/M}\,\phi_+(k,\mu) &=&
  k\theta(2w_0-k)\left[
   \int_{k/2}^{\omega_0}\!\!ds\e^{-s/M}\rho_<(s,k,\mu)+\!
  \int_{0}^{k/2}\!\!ds\e^{-s/M}\rho_>(s,k,\mu)\right]
\nonumber\\
&&{}+k\theta(k-2w_0)\int_{0}^{\omega_0}\!\!ds\,\e^{-s/M}\rho_>(s,k,\mu)
\eea
where
\bea{rho}
 \rho_<(s,k,\mu) &=& \frac{N_c}{4\pi^2}\left\{
  1+\frac{\alpha_s C_F}{2\pi}\left[\frac72+\frac{7\pi^2}{24}-
  \ln^2\frac{k}{\mu}-\frac52 \ln(x-1)-(x-1)\ln(x-1)\right.\right.
\nonumber\\&&{}-\left.\left.
  \frac12 \ln^2(x-1)-2\ln\frac{k}{\mu}[1+\ln(x-1)]+x\ln x+
   {\rm Li}_2\left(\frac{1}{1-x}\right)\right]\right\}\,,
\nonumber\\
 \rho_>(s,k,\mu) &=& \frac{\alpha_s C_F N_c}{8\pi^3}
 \left[-x+\ln(1-x)-2(1-x)\ln(1-x)+2\ln^2(1-x)
\right.\nonumber\\
&&{}\left.+2\ln\frac{k}{\mu}[x+\ln(1-x)]
 \right].
\eea
Here ${\rm Li_2}(x)$ is Euler dilogarithm function and we used a shorthand
notation $x= 2s/k$,

Neglecting $\alpha_s$ corrections for a moment, one gets a simple expression
\beq{SR3LO}
  \frac{1}{2}F^2(\mu)\e^{-\bar\Lambda/M}\,\phi_+(k,\mu) =
  \frac{N_c}{4\pi^2}\theta(2w_0-k) \, k
  \int_{k/2}^{\omega_0}\!\!ds\,\e^{-s/M}\,.
\eeq
In the local duality limit $M\to\infty$ using the sum rule expression
for $F(\mu)$ \re{SR2} with the same accuracy,
$1/2F^2(\mu) \simeq (N_c/6\pi^2)\omega_0^3$, one obtains
\beq{LOdual}
   \phi_+(k)^{\rm LD} =
    \frac{3}{4\omega_0^3}\theta(2\omega_0-k) \, k(2\omega_0-k)
\eeq
which reminds the asymptotic light-cone distribution amplitude of light mesons if
rewritten in terms of the scaling variable $\xi=k/(2\omega_0)$. For finite values
of the Borel parameter $M$ the B-meson distribution amplitude gets skewed towards
smaller values of the momentum but qualitatively remains the same, see Fig.~3.
Note that it has finite support $k<2\omega_0$ and can be interpreted as the
probability amplitude to find the light quark (on-shell) in the B-meson with
momentum $k$.

Beyond the Born approximation this simple parton-model interpretation is lost
since the distribution amplitude develops a high-momentum ``tail'' with
$k>2\omega_0$ and in this region cannot be thought of as a
probability amplitude
for the two-particle state on mass shell.
The ${\cal O}(\alpha_s)$ radiative correction turns out to be very large
($\sim$100\% of the Born term) but cancels to a large extent against the
similar large radiative correction to $F(\mu)$ \cite{BG92,BBBD92}.

The numerical results for two values of
the Borel parameter $M=0.3$~GeV and $M=0.6$~GeV are shown in Fig.~3.
%
\begin{figure}[t]
\centerline{\epsfysize5.0cm\epsffile{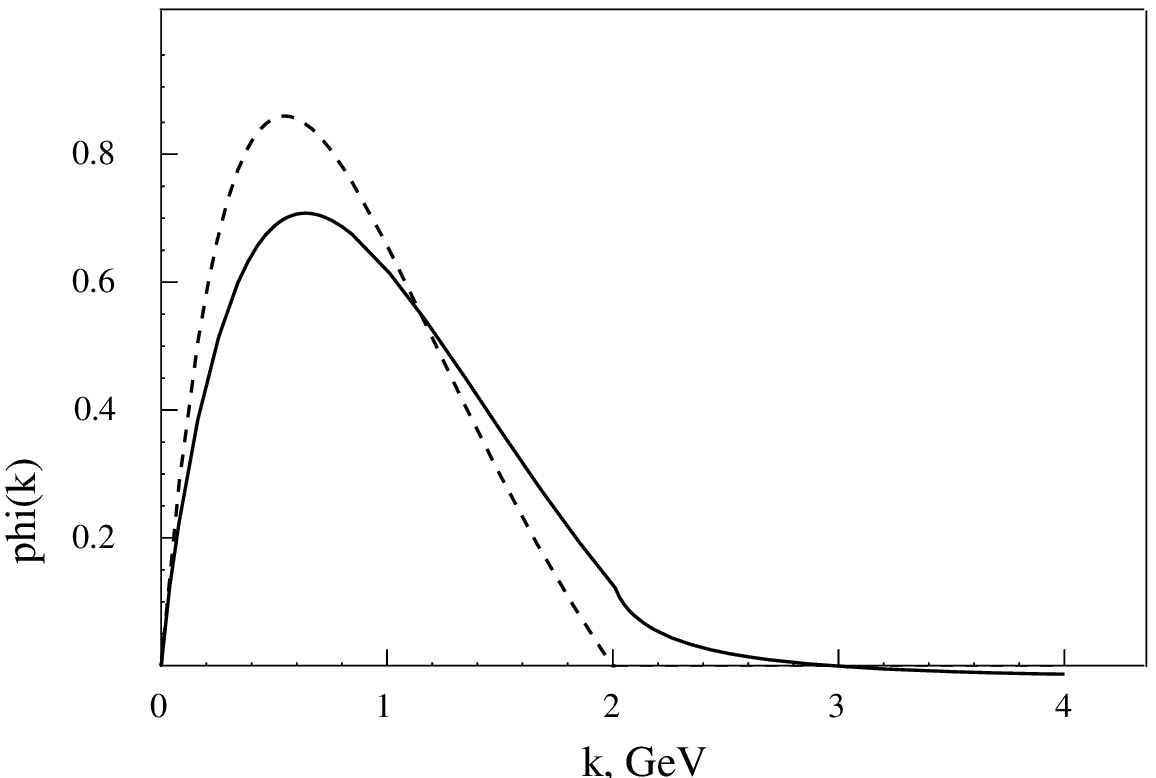}\quad
            \epsfysize5.0cm\epsffile{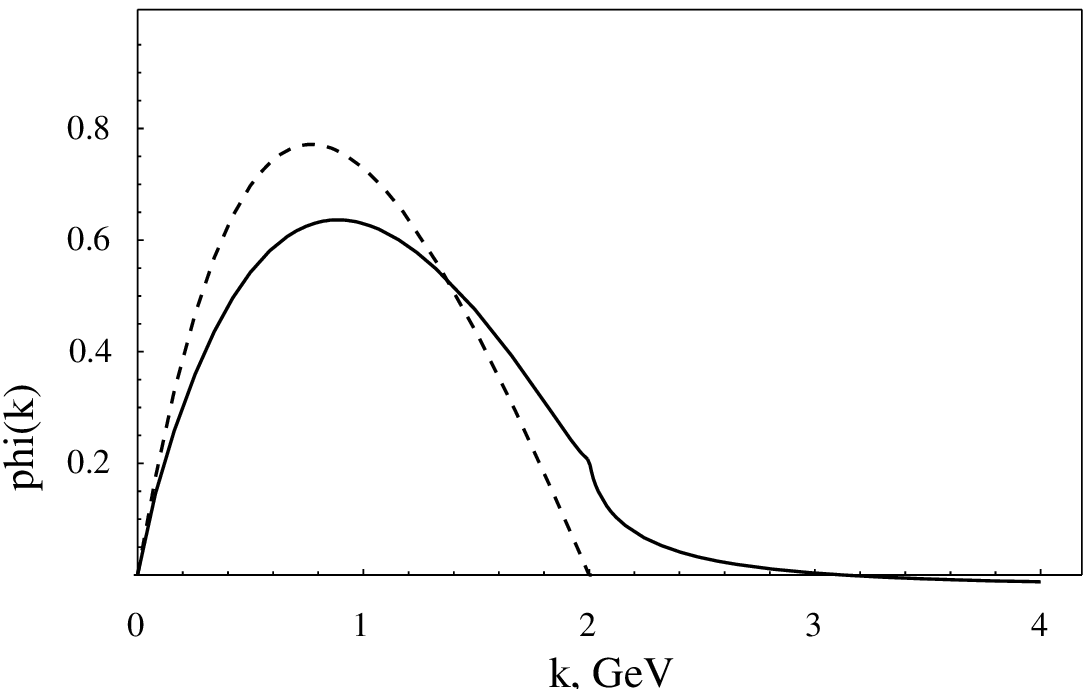}}
\caption[]{\small
B-meson distribution amplitude $\phi_+(k,\mu=1$~GeV$)$
 calculated from the sum rule
\re{SR3a} in QCD perturbation theory to leading order (dashed curves) and
next-to-leading order (solid curves) for the continuum threshold
$\omega_0$=1~GeV and two values of the Borel parameter
$M=0.3$~GeV (left panel) and $M=0.6$~GeV (right panel).
The value of the decay constant $F(\mu)$ appearing on the l.h.s.
of the sum rule is substituted by the corresponding sum rule \re{SR2}
with the appropriate accuracy (LO or NLO) and neglecting the condensate
contributions.
 }
\label{fig:3}
\end{figure}
%
We choose $w_0=1$~GeV for this plot and substitute
the coupling $F^2(\mu)$ appearing on the l.h.s. of the sum rule
\re{SR3a} by the sum rule \re{SR2} to the same accuracy, i.e.
neglecting nonperturbative corrections. In this way the dependence
on $\bar\Lambda$ cancels out and the sensitivity to other parameters
($\omega_0$ and $M$) is strongly reduced.
Indeed, it is seen in Fig.~3 that dependence
on the Borel parameter is rather mild. Note that for large $k$
the distribution amplitude becomes negative.
The asymptotic behavior is
\beq{asym}
     \phi_+(k)\sim k~~{\rm for}~~k\to0\,,\qquad
     \phi_+(k)\sim -\frac1{k}\ln(k/\mu) ~~{\rm for}~~k\gg\mu\,,
\eeq
in agreement with \cite{LN03}. Also the scale dependence of
the distribution amplitude extracted from the sum rule \re{SR3a}
agrees with \cite{LN03}. All
results are shown for $\mu=1$~GeV.

Of particular interest for the QCD description of B-decays is the
value of the first negative moment
\beq{lambdaB}
    \lambda_B^{-1}(\mu) = \int_0^\infty\frac{dk}{k}\phi_+(k,\mu)\,.
\eeq
We obtain from the sum rules:
\bea{lB10}
     \lambda_B^{-1} &=& 1.49-1.83~\mbox{\rm GeV}^{-1}\,
     \qquad \mbox{\rm for}~\omega_0=1.0~\mbox{GeV}\,,
\nonumber\\
     \lambda_B^{-1} &=& 1.79-2.08~\mbox{\rm GeV}^{-1}\,
     \qquad \mbox{\rm for}~\omega_0=0.8~\mbox{GeV}\,,
\eea
where the lower value corresponds to $M=0.6$~GeV and the higher one to
$M=0.3$~GeV for each choice of $\omega_0$. Notice that $\lambda_B^{-1}$ decreases
as $M$ increases and in the local duality limit we obtain
\beq{lB-LD}
  \lambda_B^{-1}(\mu=2\omega_0)^{\rm LD} = \frac{3}{2\omega_0}
         \left[1-\frac{\alpha_s(2\omega_0)}{\pi}
         \left(\frac53+\frac{5\pi^2}{36}
         \right)\right],
\eeq
where it is taken into account that in the limit $M\to\infty$ the sum rules
effectively become normalized at the scale $\mu=2\omega_0$ because of subtraction
of the continuum from the running coupling, cf.~\cite{BKY85}. To avoid
misunderstanding we remind that all results of this section correspond to the sum
rules in QCD perturbation theory and the given numbers will be superseded by
those in the next section where we consider the nonperturbative corrections.

\section{Sum Rules: Nonperturbative Corrections}

The primary source of nonperturbative corrections to the sum rules
in HQET is provided by the quark condensate. The corresponding diagrams
(leading and next-to-leading order) are shown in Fig.~4.
%
\begin{figure}[t]
\centerline{\epsfysize8.0cm\epsffile{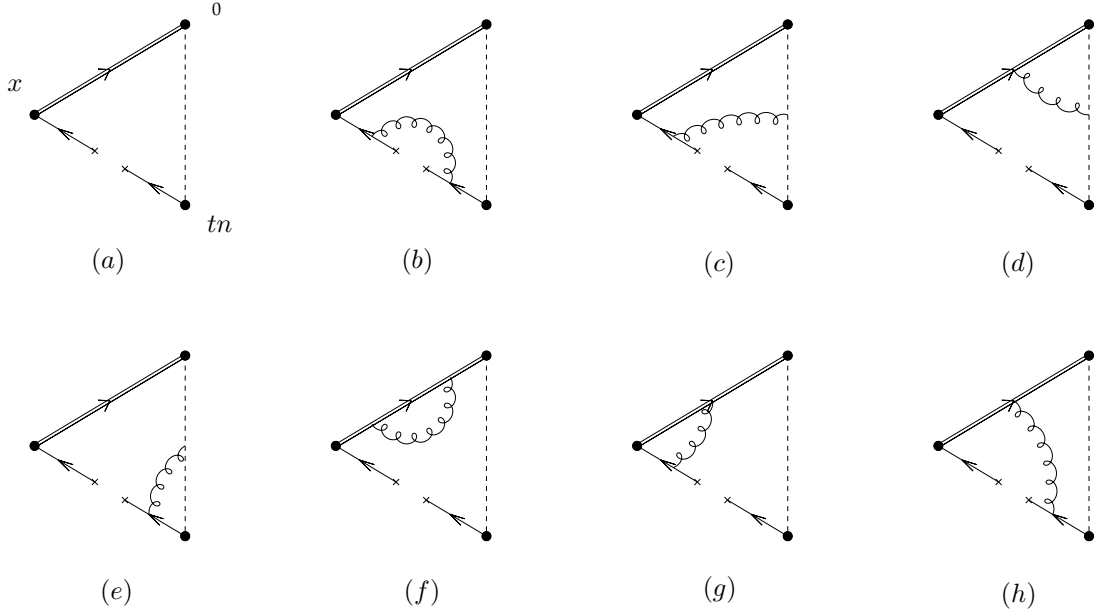}}
\caption[]{\small
Quark condensate contribution to the correlation function \re{T}.
 }
\label{fig:4}
\end{figure}
%
The leading-order contribution in Fig.~4a is simply
\beq{qqT}
  T^{\langle\bar q q\rangle}(t,\omega)
 = \frac{\langle\bar q q\rangle}{2\omega}\,.
\eeq
It does not depend on the quark-antiquark separation and gives rize to the
$\delta$-function type contribution to the r.h.s.\ of the sum rule in \re{SR3a}
\beq{qqLO}
   \ldots -\frac12 \,\langle\bar q q\rangle \delta(k)\,.
\eeq
Condensates of higher dimension produce even more singular terms, the expansion
goes in derivatives of the $\delta$-function at $k=0$. This is a well-known
problem which is familiar from the QCD sum rule studies of light-cone
distribution amplitudes of light mesons \cite{CZ} and nucleon parton
distributions \cite{BI88,BGM95}: The short-distance OPE which is the the basis of
the SVZ approach is inadequate for a calculation of distribution functions
point-by-point in the momentum fraction space. As a consequence, QCD sum rules
cannot be used for a direct calculation of distribution amplitudes (unless they
are supplemented by additional assumptions) but rather provide constraints which
have to be implemented within QCD-motivated parameterizations (models) of the
distributions, consistent with perturbative QCD. For a model to be
self-consistent, there are three conditions:
\begin{itemize}
\item{} The end-point behavior of the distributions has to be consistent
        with QCD.
\item{} The model has to be closed under the QCD evolution,
i.e. calculation of the scale dependence has to be possible and not
involve nonperturbative parameters other than those specified by the
model at the reference scale.
\item{} The model has to involve a minimum possible number of
        nonperturbative parameters.
\end{itemize}

The Chernyak-Zhitnitsky models of light-cone distribution amplitudes
of light mesons give the classical example of such an approach.
In this case one expands the distribution amplitude in a series
over orthogonal polynomials, e.g. for the pion
\beq{phipi}
   \phi_\pi(x,\mu) = 6x(1-x)\sum_{p=0,2\ldots}^\infty \varphi_p(\mu)
C^{3/2}_p(2x-1)\,,
\eeq
so that coefficients in this expansion correspond to (Gegenbauer) moments of
$\phi_\pi(x)$, and defines a {\em model} by truncating this expansion at a
certain $p=p_{\rm max}$. The first $p_{\rm max}$ coefficients are then estimated
using QCD sum rules. (In practice one takes $p_{\rm max}=2$ since estimates of
higher-order coefficients turn out to be unreliable.) The model satisfies all the
above criteria since the correct end-point behavior is built in by construction
and higher-order coefficients can only get mixed with lower-order coefficients
but not vice versa; it follows that the set of coefficients
$\{\varphi_0,\varphi_2,\ldots,\varphi_{k_{\rm max}}\}$ is closed under
renormalization\footnote{The coefficients in the Gegenbauer expansion are
renormalized multiplicatively to leading order because of conformal symmetry;
this property is, however, not essential for our argument.} and the distribution
amplitude $\phi_\pi(x,\mu)$ can be calculated at arbitrary scale from its model
at $\mu=\mu_0$. It indeed involves a minimum number of parameters, each of which
has a clear meaning in QCD as the matrix element of a certain local operator and
can eventually be calculated e.g. on the lattice.

In contrast to \re{phipi},  the B-meson distribution amplitude cannot be written
as a {\em sum} of independent terms that have autonomous QCD evolution but rather
is given by the {\em integral} in the complex moments plane \cite{LN03}. This
feature reminds evolution of parton distributions in the deep-inelastic inclusive
lepton-hadron scattering, but in difference to the latter case one cannot obtain
complex moments of the B-meson distribution amplitude by analytic continuation
from the set of real integers (as we mentioned in Sec.~2, every non-negative
moment of $\phi_+(k,\mu)$ diverges). As the result, one necessarily has a
continuous rather than discrete set of nonperturbative parameters.

One option \cite{BGM95,Grozin,GN96} is to parametrize the B-meson distribution
amplitude by the matrix element of the bilocal operator in \re{defDA} at
imaginary light-cone separation
\beq{tau}
     t = -i\tau\,,\qquad \varphi_+(\tau,\mu) = \Phi_+(-i\tau,\mu)\,.
\eeq
Obviously
\beq{fur}
   \varphi_+(\tau,\mu) = \int_0^\infty\! dk\,\e^{-\tau k}\phi_+(k,\mu)
\eeq
and the parameter $\lambda_B$ is given by the simple integral
\beq{lamB}
   \lambda_B^{-1}(\mu) = \int_0^\infty d\tau\,\varphi_+(\tau,\mu)\,.
\eeq
The purpose of going over to imaginary light-cone times (distances) is similar to
that of the usual Wick rotation: In this way the oscillating exponents
corresponding to the light-cone time dependence of intermediate states
propagating along the light-cone are converted to falling exponents suppressed by
the energy of the state, where the light-cone quantization is implied.
Simultaneously, the normalization scale $\mu$ acquires the physical meaning of
the cutoff in energy of the intermediate states. Note that the renormalization of
$\varphi_+(\tau,\mu)$ only involves the distribution at smaller light-cone
separations,  cf. \re{Zcoord}. This implies that knowledge of
$\varphi_+(\tau,\mu)$ at small distances up to  $\tau<\tau_{\rm max}$ is
sufficient to calculate its scale dependence in the same distance range, in
agreement with the self-consistency criterium formulated above.

On the other hand, it is easy to understand that the function
$\varphi_+(\tau,\mu)$ can be calculated at small $\tau$ using  OPE; expansion in
vacuum condensates of increasing dimension corresponds to the Laurent expansion
of $\varphi_+(\tau,\mu)$ in powers of $\tau$, which is modified by calculable
perturbative corrections. The condensate expansion seems to be under control up
to distances of order $\tau\sim 1$~GeV$^{-1}$ (that is, of order 0.2~fm). At
larger distances the OPE diverges and one has to either truncate
$\varphi_+(\tau,\mu)$ at a certain $\tau_{\rm max}$ or rely on a certain model
for the large $\tau$ behaviour. Provided that the nonperturbative corrections
decrease sufficiently fast for large $\tau$ one can hope that the model
assumptions do not lead to a large uncertainty in the overall result.

To illustrate this construction,  we have calculated the quark condensate
contribution including the $\alpha_s$-correction, see Fig.~4, the
contribution of the gluon condensate, Fig.~5, and the contribution
%
\begin{figure}[t]
\centerline{\epsfysize4.0cm\epsffile{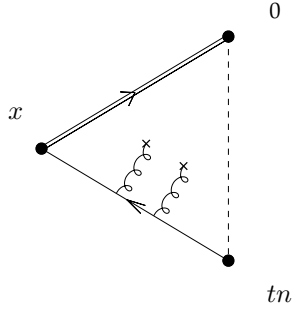}}
\caption[]{\small
Gluon condensate contribution to the correlation function \re{T}.
Only this diagram contributes in the Fock-Schwinger gauge.}
\label{fig:5}
\end{figure}
%
of the mixed condensate $\langle \bar q \sigma gG q\rangle
\simeq m_0^2 \langle\bar q q\rangle $ which is obtained as the
expansion of the diagram in Fig.~4a in the background gluon field.
The resulting sum rule in which we have also included the perturbative
contribution of Fig.~2 reads
\bea{SR3bac}
\lefteqn{
   \frac{1}{2}F^2(\mu)\e^{-\bar\Lambda/M}\,\varphi_+(\tau,\mu) \,=}
 \\&=&
   \int_{0}^{\omega_0}\!\!ds\,\e^{-s/M}\rho_{\rm pert}(s,\tau,\mu)
    -\frac12\langle\bar q q\rangle\Bigg\{1+\frac{\alpha_sC_F}{2\pi}
 \Big[3-\frac{5\pi^2}{24}-\ln^2(\tau\mu \e^{\gamma_E})
- \ln(\tau\mu \e^{\gamma_E}) \nonumber\\ &&{} -\ln(1+2\tau M) -{\rm L}_2(-2\tau
M)\Big]\Bigg\}
 +\frac{1}{48}\left\langle\frac{\alpha_s}{\pi}G^2\right\rangle
\frac{M\tau^2}{(1+2\tau M)^2}
+\frac{1}{32}\frac{m_0^2}{M^2}\langle\bar q q\rangle
   \left(1+2\tau M\right)\,,
\nonumber
\eea
where the perturbative spectral density can be read off Eq.~\re{SR3a}. The
contribution of the gluon condensate turns out to be very small and will be
neglected in what follows. We further note that the sum of the diagram in Fig.~4d
and one half of the heavy quark self-energy correction in Fig.~4f define the
universal renormalization factor of the Wilson line built of the light-like
segment of length $-i\tau$ and the time-like segment of length $-i/M$. It can be
shown that the corresponding contributions $\sim (\alpha_sC_F)^n$ exponentiate to
all orders \cite{Gatheral} and produce a Sudakov-like exponential suppression
factor
\beq{sud}
   S(\tau,M,\mu) =
   \exp\left\{-\frac{\alpha_sC_F}{2\pi}\left[\ln^2(\tau\mu \e^{\gamma_E})
+ \frac{5\pi^2}{24} -1 -\ln \frac{\mu \e^{\gamma_E}}{2M}+ {\rm L}_2(-2\tau M)
  \right]\right\}
\eeq
which is the same for the quark and the quark-gluon condensate.
Note that ${\rm L}_2(-2\tau M) \sim -\frac12 \ln^2(2\tau M)$ for
$\tau \gg 1/M$.
We end up with an improved sum rule
\bea{SR3b}
\lefteqn{
   \frac{1}{2}F^2(\mu)\e^{-\bar\Lambda/M}\,\varphi_+(\tau,\mu) \,=}
 \\&=&
   \int_{0}^{\omega_0}\!\!ds\,\e^{-s/M}\rho_{\rm pert}(s,\tau,\mu)
    -\frac12\langle\bar q q\rangle\,S(\tau,M,\mu)\,
\Bigg\{1+\frac{\alpha_sC_F}{2\pi}
 \Big[2
- \ln(\tau\mu \e^{\gamma_E}) - \ln \frac{\mu \e^{\gamma_E}}{2M} \nonumber\\ &&{}
-\ln(1+2\tau M)\Big]
-\frac{1}{16}\frac{m_0^2}{M^2}\left(1+2\tau M\right)\Bigg\}
\nonumber
\eea
in which the double-logarithmic corrections  to the quark and
quark-gluon chiral condensates are resummed. We do not attempt the
similar resummation in the perturbative contribution since its effect
is negligible compared to the $1/\tau^2$ falloff inherited from the
Born term.

%
\begin{figure}[t]
\centerline{\epsfysize4.4cm\epsffile{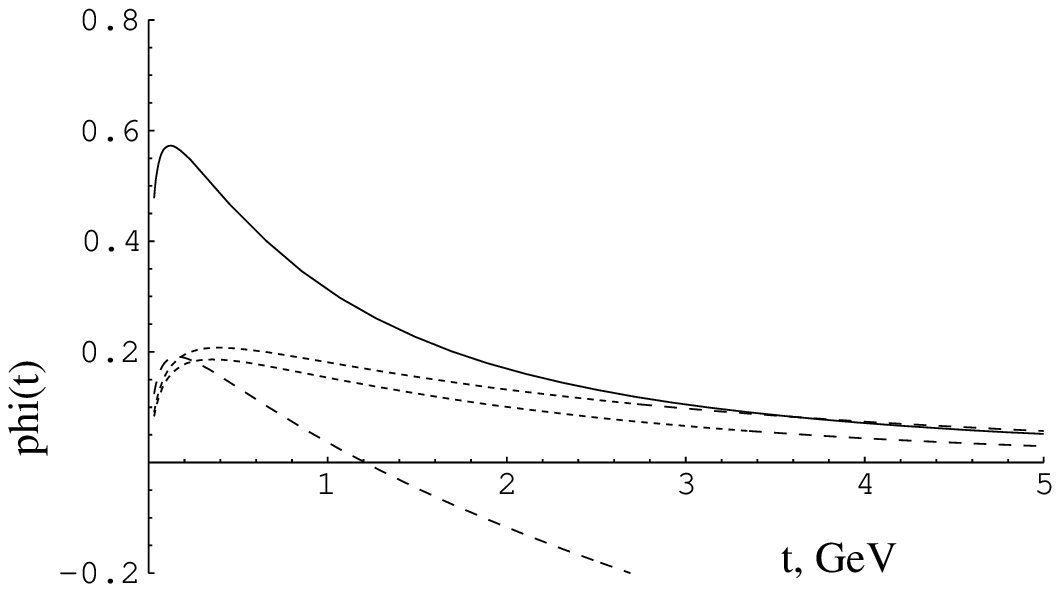}\quad
            \epsfysize4.4cm\epsffile{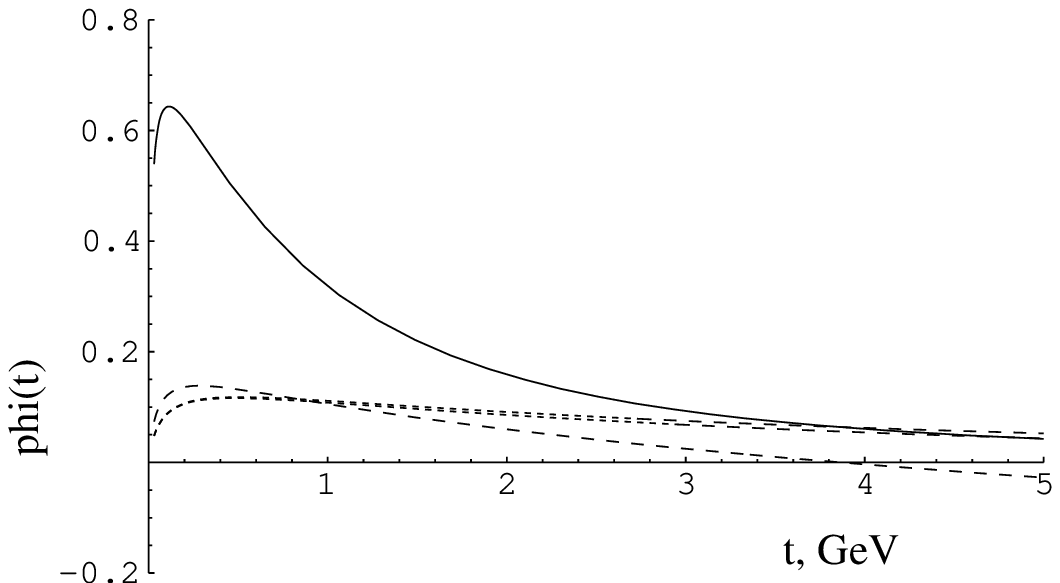}}
\caption[]{\small Perturbative (solid curves) and nonperturbative (long dashes)
contributions to the B-meson distribution amplitude $\varphi_+(\tau,\mu=1$~GeV$)$
calculated from the sum rule \re{SR3b} to the NLO accuracy. In addition, the
nonlocal condensate models \re{model1}, \re{model2} of resummed nonperturbative
contributions to the sum rule, cf. \re{nonphi}, are shown by short dashes. The
continuum threshold is chosen to be $\omega_0$=1~GeV and two values of the Borel
parameter are used: $M=0.3$~GeV (left panel) and $M=0.6$~GeV (right panel). The
value of the decay constant $F(\mu)$ appearing on the l.h.s. of \re{SR3b} is
substituted by the corresponding sum rule \re{SR2}.
 }
\label{fig:6}
\end{figure}
%
%
\begin{figure}[t]
\centerline{\epsfysize4.4cm\epsffile{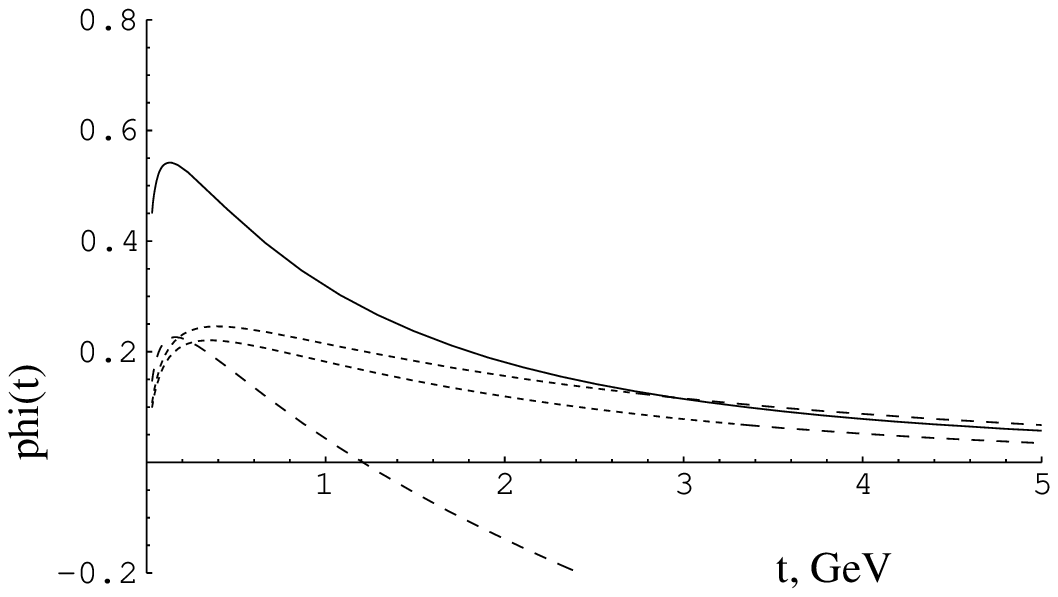}\quad
            \epsfysize4.4cm\epsffile{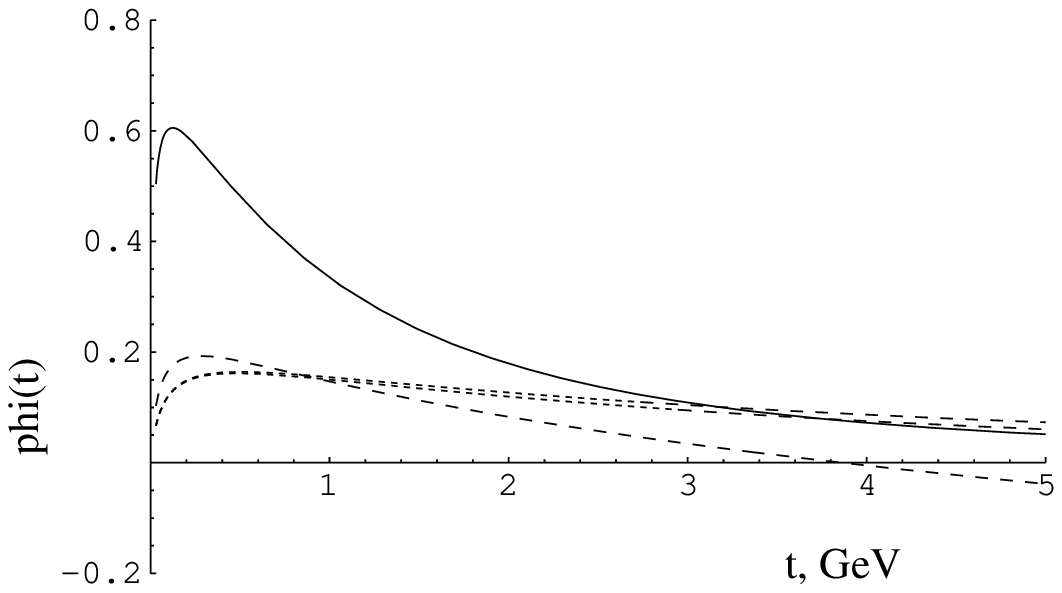}}
\caption[]{\small
Same as in Fig.~6, with a different value of the
continuum threshold $\omega_0=0.8$~GeV.
 }
\label{fig:7}
\end{figure}
%

The perturbative and the nonperturbative contributions to the sum rule result
\re{SR3b} for $\varphi_+(\tau,\mu)$ are shown separately as a function of
distance $\tau$ in Fig.~6 ($\omega_0=1$~GeV) and Fig.~7 ($\omega_0=0.8$~GeV) for
two different values of the Borel parameter $M=0.3$~GeV and $M=0.6$~GeV. Note
that at small distances the nonperturbative corrections are significantly smaller
than the perturbative contribution. The nonperturbative correction turns to zero
at a certain value of $\tau$ as the result of the cancellation between the quark
condensate contribution ($\sim$~{\em const}) and that of the mixed condensate
($\sim \tau$). This cancellation of course cannot be taken seriously and only
indicates that the OPE breaks down since the hierarchy of contributions is lost.
We conclude that the classical QCD sum rule is only valid up to light-cone
distances of order $1-3$~GeV$^{-1}$, depending on the value of the Borel
parameter. A rough estimate for the nonperturbative contribution to
$\lambda_B^{-1}$ in the strict OPE-based approach is, therefore, given by the
integral over the region of small $\tau$ where the correction is still positive,
that is up to the crossing point with the zero axis. In order to get an estimate
of a possible nonperturbative contribution from large distances we use the
concept of a nonlocal quark condensate introduced in \cite{MR88} and later used
rather extensively in QCD sum rule calculations of the distribution amplitudes of
light mesons by the Dubna group \cite{nonloc1,BM02}. The same approach was taken
up in \cite{GN96}.

The nonlocal quark condensate presents a model for a partial
resummation of the OPE to all orders in terms of the vacuum
expectation value of the single nonlocal
operator
\beq{NLC}
   \langle 0|\bar q(x)[x,0]q(0)|0\rangle =
    \langle \bar q q\rangle \int_0^\infty\!d\nu\, \e^{\nu x^2/4} f(\nu)\,.
\eeq
The first two moments of $f(\nu)$ are fixed by the OPE:
\beq{nlcmom}
   \int_0^\infty\!d\nu\, f(\nu) =1\,,\qquad
   \int_0^\infty\!d\nu\, \nu f(\nu) = \frac14 m_0^2\,,
\eeq
and in addition one requires that the correlation function \re{NLC}
decreases exponentially at large Euclidian separations $x^2\to-\infty$.
The two simplest choices are
\beq{model1}
     \mbox{Model I}:~~f(\nu) = \delta(\nu-m_0^2/4)~~\cite{MR88}
\eeq
corresponding to the Gaussian large-distance behavior $\sim \exp[-|x^2|m_0^2/16]$
and
\beq{model2}
     \mbox{Model II}:~~f(\nu) =
 \frac{\lambda^{p-2}}{\Gamma(p-2)} \nu^{1-p} \e^{-\lambda/\nu}\,,
  \quad p= 3+\frac{4\lambda}{m_0^2}~~\cite{BGM95}
\eeq
corresponding to  $\langle 0|\bar q(x)q(0)|0\rangle \sim
\exp[-\lambda\sqrt{-x^2}]$. Here $\lambda$ is a parameter with
physical meaning of the vacuum quark correlation length.
In this work we take $\lambda=400$~MeV as a representative number,
cf \cite{BM02,Radyushkin91}.
We will see that sensitivity of the sum rules to the shape of
 $f(\nu)$ is in fact small; the major shortcoming of this approach
is rather that other condensates (e.g. the nonlocal
quark-antiquark-gluon condensate) are not included and there exists no
parameter that would justify such a truncation.

Nonlocality of the quark condensate is  easy to implement within
our sum rules and it amounts to a simple substitution in Eq.~\re{SR3b}
(cf. \cite{GN96})
\bea{nonphi}
&& \hspace*{-20mm}-\frac12\langle\bar q q\rangle\,S(\tau,M,\mu) \Bigg\{1+{\cal
O}(\alpha_s)- \frac{1}{16}\frac{m_0^2}{M^2}\left(1+2\tau M\right)\Bigg\}
\nonumber \\&\to& -\frac12\langle\bar q q\rangle\,S(\tau,M,\mu)
 \int_0^\infty\!d\nu\, f(\nu)\, \e^{-\nu(1+2\tau M)/(4M^2)}.
\eea
Note that the mixed condensate contribution is now included as a part of the
nonlocal condensate and we neglect the $\alpha_s$-correction to the (local) quark
condensate (but retain the Sudakov exponent). The results are shown in Fig.~6 and
Fig.~7 by short dashes; the lower and the higher of the curves correspond to the
choice in \re{model1} and \re{model2}, respectively. The corresponding results
for $\lambda_B^{-1}$ are, for $\mu=1$~GeV:
\bea{l10}
 \omega_0 = 1~\mbox{\rm GeV}\,,\phantom{0.}\quad M= 0.6~\mbox{\rm GeV}:&\qquad&
 \lambda_B^{-1} = 1.23 +
 \left\{\begin{array}{c}0.26 \\ 0.60 \\ 0.83\end{array}\right.
 = 1.95\pm 0.23~\mbox{GeV}^{-1}\,,
\nonumber\\
 \omega_0 = 1~\mbox{\rm GeV}\,,\phantom{0.}\quad M= 0.3~\mbox{\rm GeV}:&\qquad&
 \lambda_B^{-1} = 1.32 +
 \left\{\begin{array}{c}0.13 \\ 0.54 \\ 0.88\end{array}\right.
 = 2.03\pm 0.29~\mbox{GeV}^{-1}\,,
\nonumber\\
 \omega_0 = 0.8~\mbox{\rm GeV}\,,\quad M= 0.6~\mbox{\rm GeV}:&\qquad&
 \lambda_B^{-1} = 1.36 +
 \left\{\begin{array}{c}0.35 \\ 0.84 \\ 1.16\end{array}\right.
 = 2.36\pm 0.33~\mbox{GeV}^{-1}\,,
\nonumber\\
 \omega_0 = 0.8~\mbox{\rm GeV}\,,\quad M= 0.3~\mbox{\rm GeV}:&\qquad&
 \lambda_B^{-1} = 1.39 +
 \left\{\begin{array}{c}0.15 \\ 0.64 \\ 1.05\end{array}\right.
 = 2.24\pm 0.35~\mbox{GeV}^{-1},
\eea
where the first number gives the perturbative contribution
(the difference to \re{lB10} is due to the different value used
 for $F(\mu)$) and the three numbers under the brace correspond to
three different estimates for the nonperturbative contribution: 1) quark and
mixed condensate contribution to \re{SR3b} restricted to the positivity region,
2) nonlocal condensate model I \re{model1} and 3) nonlocal condensate   model II
\re{model2}. The first (upper) number should be considered as an estimate of the
nonperturbative correction from below while the difference between the two lower
ones characterizes the uncertainty in the choice of the parametrization of the
nonlocal condensate.  We take the average between the two models as our central
value, and one half of the difference between this central value and the first
(upper) number, coming from local OPE, as an estimate of the overall uncertainty
of the result. In other words, we ascribe 50\% uncertainty to the extrapolation
of the nonperturbative contribution to large distances as suggested by the
nonlocal condensate model, which is rather conservative, cf.~\cite{nonloc1}. From
the numbers in Eq.~\re{l10} we obtain the final result
\beq{otvet1}
      \lambda_B^{-1}(\mu=1~\mbox{\rm GeV}) = 2.15\pm 0.5 ~\mbox{GeV}^{-1}
\eeq
or
\beq{otvet2}
      \lambda_B(\mu=1~\mbox{\rm GeV}) = 460 \pm 110~ ~\mbox{MeV}\,.
\eeq
Our value of $\lambda_B$ is somewhat larger than the number accepted in
\cite{BBNS,DS03} $\lambda_B=0.35\pm 0.15$~GeV, although consistent with
it within errors. It is also consistent with the rough estimate
$\lambda_B\simeq 0.6$~GeV derived in \cite{BK03}.

As follows from \re{defZ} and \re{Zmom}, the scale dependence of
$\lambda_B$ also involves the first logarithmic moment of the
distribution amplitude \cite{LN03,Ball03} \beq{renlam}
  \lambda_B^{-1}(\mu) =
    \left[1+\frac{\alpha_sC_F}{2\pi}\ln\frac{\mu}{\mu_0}\right]\,
     \lambda_B^{-1}(\mu_0)
     -\frac{\alpha_sC_F}{\pi}\ln\frac{\mu}{\mu_0}
      \int_0^\infty \frac{dk}{k}\ln\frac{\mu_0}{k}\phi_+(k,\mu_0)\,,
\eeq
where $({\alpha_sC_F}/{\pi})\ln({\mu}/{\mu_0})< 1$. We define
\beq{sB}
 \sigma_B(\mu) =
   \lambda_B(\mu) \int_0^\infty
\frac{dk}{k}\ln\frac{\mu}{k}\,\phi_+(k,\mu)
   = \lambda_B(\mu) \int_0^\infty\!d\tau\,\ln(\tau \mu \e^{\gamma_E})
      \,\varphi_+(\tau,\mu)
\eeq
and calculate $\sigma_B(1~\mbox{\rm GeV})$ from the QCD sum rule
\re{SR3b} repeating the same procedure as explained above for
$\lambda_B$. Without going into details we simply quote the final result
\beq{SBres}
  \sigma_B(\mu=1~\mbox{\rm GeV}) = 1.4 \pm 0.4\,.
\eeq
Note that $\sigma_B(\mu)$ defines the average value of $\ln \mu/k$ in
the integral for the first inverse moment $\lambda_B^{-1}$,
so that the number in \re{SBres} implies that main contribution to
$\lambda_B^{-1}$ comes from momenta $\sim 250$~MeV.
With this value for $\sigma_B$,
the two contributions ${\cal O}(\alpha_s)$ in \re{renlam}
tend to cancel each other to a large extent, so that the remaining
scale dependence of $\lambda_B^{-1}$ is weak.

A simple model of the distribution amplitude $\phi_+(k,\mu)$
with given values of the parameters $\lambda_B$ and $\sigma_B$ and
correct asymptotic behavior can be chosen as
\beq{phimod1}
   \phi_+(k,\mu=1~\mbox{\rm GeV}) =
\frac{4\lambda_B^{-1}}{\pi}\frac{k}{k^2+1}
\left[\frac{1}{k^2+1} -\frac{2(\sigma_B-1)}{\pi^2}\ln k\right]
\eeq
($k$ in units of GeV).
Using the values of $\lambda_B$ and $\sigma_B$ in \re{otvet1},
\re{SBres} one obtains the distribution shown in Fig.~8 by the
solid curve. For comparison, we also show in this plot a typical
distribution amplitude obtained from QCD sum rules in perturbation
theory \re{SR3a} in Sec.~3, cf. Fig.~3.
The effect of nonperturbative corrections is to shift the distribution
towards softer momenta, which is natural.
%
\begin{figure}[t]
\centerline{\epsfysize5.0cm\epsffile{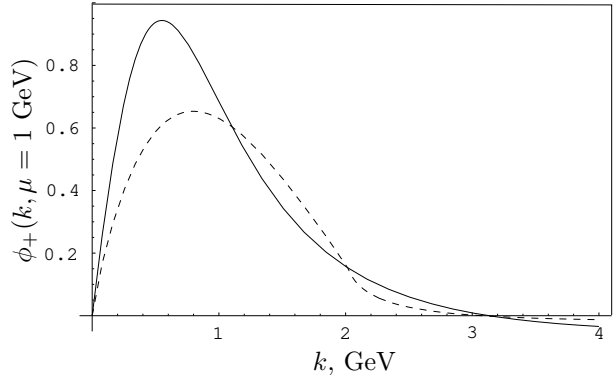}
            }
\caption[]{\small A QCD model for the B-meson distribution amplitude \re{phimod1}
(solid curve) compared with the perturbative sum rule prediction \re{SR3a}
 with M=0.45~GeV, $\omega_0=1$~GeV (dashed curve).
 }
\label{fig:8}
\end{figure}
%
One minor drawback of such a parametrization is that the set of parameters
$\lambda_B$ and $\sigma_B$ is not closed under renormalization.
In view of a very limited range of scales that are interesting for
B-decay phenomenology this seems to be not a  problem, however.

To summarize, in this paper we have derived QCD sum rules for the
B-meson distribution amplitude \re{defDA} and, in particular, obtained
an estimate of its first inverse moment $\lambda_B^{-1}$ \re{otvet1}
and the parameter $\sigma_B$ \re{sB} that characterizes the shape
of the distribution, see \re{SBres}. A simple model is suggested \re{phimod1}
that incorporates all existing constraints. We believe that our estimates
are interesting for the studies of the heavy quark limit in exclusive
B-decays and can be used in a broad context.
Concrete applications go beyond the task of this work.

\section*{Acknowledgements}
V.B. is grateful to the Laboratoire de Physique Th\'eorique,
Universite Paris XI, for hospitality during his sabbatical stay in
Orsay and CNRS for the financial support. Work of D.I. is
partially supported by Alexander von Humboldt Foundation and by
grants RFBR 02-02-17884 and INTAS 00-00679.

\end{document}